\documentclass{article}
\usepackage[utf8]{inputenc}
\usepackage[margin=1.0in]{geometry}
\usepackage{hyperref}
\usepackage{amsmath, bbm}

\title{A Note On Interpreting Canary Exposure}
\author{Matthew Jagielski, Google DeepMind}
\date{\vspace{-5mm}}

\usepackage[numbers]{natbib}
\usepackage{graphicx}

\begin{document}

\maketitle
\begin{abstract}
    Canary exposure, introduced in Carlini et al.~\cite{carlini2019secret}
    is frequently used to empirically evaluate, or audit, the privacy of machine learning model training. The goal of this note is to provide some intuition on how to interpret canary exposure, including by relating it to membership inference attacks~\cite{homer2008resolving, shokri2017membership} and differential privacy~\cite{dwork2006calibrating}.
\end{abstract}

\section{Background}

\textbf{Secret Sharer}~\cite{carlini2019secret} is a privacy evaluation framework that injects secret \emph{canaries} into model training and attempts to extract them from a trained model $f$. This framework is frequently used when model training is very expensive~\cite{huang2022detecting, malek2021antipodes, ramaswamy2020training, stock2022defending, thakkar2020understanding, zanella2020analyzing, zhang2021wide}, and is also used as a proxy for understanding harmful memorization risk~\cite{anil2023palm, tramer2022truth}. This evaluation requires first selecting a distribution over examples (e.g. uniformly random strings), from which $m$ ``canaries'' $C = \lbrace c_i\rbrace_{i=1}^m$ are sampled and added into the model's training set, and $n$ ``references'' $R = \lbrace r_i\rbrace_{i=1}^n$ are sampled and withheld for comparison. Carlini et al.~\cite{carlini2019secret} propose measuring memorization with the exposure metric, which can be calculated for each canary $c_i$ by computing the rank of its loss, $\ell(c_i)$, among the losses of all references $r_i$, as:
$$
\text{Exposure}(c_i) = \log_2(n) - \log_2(\text{Rank}(\ell(c_i), \lbrace \ell(r_i) \rbrace_{i=1}^n)).
$$
Here, Rank gives a value between 1 ($c_i$'s loss is lower than all references), and $n+1$ ($c_i$'s loss is higher than all references).
Intuitively, a canary is maximally \emph{exposed} by the model when it is the most probable example, and has a rank of 1, or an exposure of $\log_2(n)$. If no memorization occurs, the rank will be $1+n/2$ on average, where exposure is about 1. Aside from these extremes, exposure can be confusing, especially when aggregating exposures across multiple canaries; this note gives a couple tools for thinking about exposure, by relating it to membership inference and differential privacy.

\textbf{Membership Inference}~\cite{homer2008resolving, shokri2017membership} is a privacy attack whose goal is to determine whether a given target record $z$ was contained in the training set of a model $f$. Historically, the membership inference attack literature has measured attack success with attack accuracy, but recently this has been replaced by measuring the attack's True Positive Rate (TPR) at a small False Positive Rate (FPR)~\cite{carlini2022membership, micochallenge}.

\textbf{Differential Privacy} (DP)~\cite{dwork2006calibrating} is a formal notion of data privacy, which, in particular (and all we need in this note), bounds the success rate of membership inference attacks~\cite{kairouz2015composition}. DP typically uses a parameter $\varepsilon$, where a smaller $\varepsilon$ provides stronger privacy, and permits less successful membership inference attacks: an algorithm satisfying $\varepsilon$-DP bounds the success of any membership inference attack as $\tfrac{\text{TPR}}{\text{FPR}}\le \exp(\varepsilon)$.

\section{Interpreting Exposure}

\subsection{Random Guessing Baseline}
It's always good to know what your baselines are when you interpret results. In membership inference experiments, it is common to use a ``random guessing'' baseline (usually set up to have 50\% accuracy). Random guessing is particularly useful for privacy attacks, as a very bad attack will not perform much better than random guessing, and a very good defense will make every attack perform as well as random guessing. For canary exposure, such a baseline can be constructed by using an uninformative loss function that results in a random ordering of all canaries. When you're running your own canary experiments, it will be helpful to compare these metrics against those obtained with random guessing.

Let's work out what some natural baselines should look like for exposure.
A common one is the expected exposure of a canary: with a bit of work, we can compute this to be roughly $1/\ln(2)\approx1.44$.\footnote{Start by writing out average exposure as $\tfrac{1}{n+1}\sum_{i=1}^{n+1} (\log_2(n)-\log_2(i))$. We can expand this out to $\log_2(n)-\tfrac{1}{n+1}\log_2((n+1)!)$. Using Stirling's approximation to $n!$ and simplifying gives roughly $1/\ln(2)$.}
This deserves a second of pause---the expected \emph{rank} of a canary is $1+n/2$, a rank which gives an exposure of $1$ as $n$ gets large, but the logarithm in the exposure formula biases it above 1.
I personally prefer looking at the \emph{median} exposure, as it is less sensitive to the outliers that are common in canary experiments.
When $m$ is large, the median exposure also, quite nicely, has an expected value of 1 with a random guessing baseline.\footnote{The relevant tool here is \emph{order statistics}. For example, asymptotically, the expected value of the median of a sample is the median of the distribution; see e.g. \url{https://en.wikipedia.org/wiki/Order_statistic\#Large_sample_sizes}.} Other quantiles are similarly easy to analyze, and may be of interest for measuring the risk to outliers; the random guessing baseline gives an exposure of 2 for the 75th percentile of canaries (the most exposed $m/4$ canaries).

\subsection{Exposure, Differential Privacy, and Membership Inference}
Here, we're going to see how looking at the median also helps us connect canary exposure to DP and membership inference. First, let's write the median canary loss as $\ell_{\text{Median}}$, and the canary with that loss as $c_{\text{Median}}$. Actually, replacing $\ell_{\text{Median}}$ with any loss value will work for our calculations here, but the median will make calculations simple. We now consider the following two probabilities:
$$
p_c = \text{Pr}_{c\sim C}[\ell(c) < \ell_{\text{Median}}],~~~ p_r = \text{Pr}_{r\sim R}[\ell(r) < \ell_{\text{Median}}].
$$

Now, $p_c$ is the probability a canary will have a loss smaller than $\ell_{\text{Median}}$, but we've chosen $\ell_{\text{Median}}$ to be the median, so $p_c=0.5$ by definition. Similarly, $p_r$ is the probability that a reference example will have a loss smaller than $\ell_{\text{Median}}$, which is equivalent to $\tfrac{1}{n}\text{Rank}(\ell_{\text{Median}}, \lbrace \ell(r_i) \rbrace_{i=1}^n)-\tfrac{1}{n}$, since Rank counts the references with smaller loss and adds 1. A short calculation gives us that the exposure of the median loss canary is roughly $\text{Exposure}(c_{\text{Median}})=\log_2(1/p_r)$ (ignoring the small $\tfrac{1}{n}$ subtracted from Rank).

Another way of thinking of $p_c$ and $p_r$ is actually as the TPR and FPR of a membership inference attack. The membership inference attack here classifies an example $z$ as IN the training set if $\ell(z) < \ell_{\text{Median}}$. The TPR is the proportion of canaries (the ``members'') which are guessed as IN, which is exactly $p_c$. The FPR is the proportion of references (the ``nonmembers'') which are guessed as IN, which is $p_r$. Here is our connection between exposure and membership inference. Now, recalling that DP bounds membership inference with $\tfrac{\text{TPR}}{\text{FPR}}\le \exp(\varepsilon)$, we can convert the median exposure to a lower bound on $\varepsilon$. To do so, we write 
$$
\varepsilon \ge \ln\left(\frac{\text{TPR}}{\text{FPR}}\right) = \ln\left(\frac{p_c}{p_r}\right) = \ln\left(\frac{1}{2p_r}\right),
$$
with the final step using that $p_c=0.5$. Now, recalling that $\text{Exposure}(c_{\text{Median}})=\log_2(1/p_r)$, we can write our lower bound as
$$\varepsilon\ge \ln(2)\left(\text{Exposure}(c_{\text{Median}})-1\right).$$

Using other loss thresholds instead of $\ell_{\text{Median}}$ result in similar lower bounds. By looking at lower FPRs, for example, we can evaluate the privacy risk to more outlier examples.

Some caveats:
\begin{enumerate}
    \item The above is implied by existing bounds on reconstruction. It is a special case of the \emph{probability preservation property} introduced in Stock et al.~\cite{stock2022defending}, which uses DP to analyze the change in probabilities of model generations when changing any training example, including canaries. Corollary 3 of Balle et al.~\cite{balle2022reconstructing} also implies our derivation. 
    \item The probabilities $p_c$ and $p_r$ above are estimated from samples, rather than being the true population values. To convert them to a valid lower bound on the $\varepsilon$ parameter, as is done in privacy auditing~\cite{jagielski2020auditing}, it is important to use statistical techniques \cite{jagielski2020auditing, lu2022general, zanella2022bayesian} to correct for sampling error. These probabilities also assume that all canary and reference losses are independent, which can be assumed as a heuristic~\cite{zanella2022bayesian}, or addressed with very recent analysis that avoids this assumption (Theorem 2.1 in Steinke et al.~\cite{steinke2023privacy} or Algorithm 1 in Pillutla et al.~\cite{pillutla2023unleashing}).
    \item It is common to duplicate canaries multiple times in the training set, to make it easier to measure exposure. To appropriately adjust the resulting $\varepsilon$, the \emph{group privacy} property of DP should be used, by dividing the obtained $\varepsilon$ by the number of duplicates (when using the standard DP definition).
\end{enumerate}

Final suggestions:
\begin{enumerate}
    \item When aggregating $m$ canary exposures, compare to random guessing baselines.
    \item When measuring memorization risk with canary exposure, report a corresponding $\varepsilon$ lower bound, at a low FPR if $m$ is large enough.
    \item Calibration, as introduced in Watson et al.~\cite{watson2021importance} and adapted to canary extraction in Tram\`{e}r et al.~\cite{tramer2022truth}, should be used when possible, as it can significantly increase exposure.
\end{enumerate}

\section*{Acknowledgements}
Many thanks to Georgie Evans, Jamie Hayes, Chenyue Lu, Thomas Steinke, Andreas Terzis, and Om Thakkar, who have all provided helpful feedback.

\bibliographystyle{plain}
\bibliography{main}

\begin{thebibliography}{10}

\bibitem{anil2023palm}
Rohan Anil, Andrew~M Dai, Orhan Firat, Melvin Johnson, Dmitry Lepikhin,
  Alexandre Passos, Siamak Shakeri, Emanuel Taropa, Paige Bailey, Zhifeng Chen,
  et~al.
\newblock Palm 2 technical report.
\newblock {\em arXiv preprint arXiv:2305.10403}, 2023.

\bibitem{balle2022reconstructing}
Borja Balle, Giovanni Cherubin, and Jamie Hayes.
\newblock Reconstructing training data with informed adversaries.
\newblock In {\em 2022 IEEE Symposium on Security and Privacy (SP)}, pages
  1138--1156. IEEE, 2022.

\bibitem{carlini2022membership}
Nicholas Carlini, Steve Chien, Milad Nasr, Shuang Song, Andreas Terzis, and
  Florian Tramer.
\newblock Membership inference attacks from first principles.
\newblock In {\em 2022 IEEE Symposium on Security and Privacy (SP)}, pages
  1897--1914. IEEE, 2022.

\bibitem{carlini2019secret}
Nicholas Carlini, Chang Liu, {\'U}lfar Erlingsson, Jernej Kos, and Dawn Song.
\newblock The secret sharer: Evaluating and testing unintended memorization in
  neural networks.
\newblock In {\em USENIX Security Symposium}, volume 267, 2019.

\bibitem{dwork2006calibrating}
Cynthia Dwork, Frank McSherry, Kobbi Nissim, and Adam Smith.
\newblock Calibrating noise to sensitivity in private data analysis.
\newblock In {\em Theory of Cryptography: Third Theory of Cryptography
  Conference, TCC 2006, New York, NY, USA, March 4-7, 2006. Proceedings 3},
  pages 265--284. Springer, 2006.

\bibitem{homer2008resolving}
Nils Homer, Szabolcs Szelinger, Margot Redman, David Duggan, Waibhav Tembe,
  Jill Muehling, John~V Pearson, Dietrich~A Stephan, Stanley~F Nelson, and
  David~W Craig.
\newblock Resolving individuals contributing trace amounts of dna to highly
  complex mixtures using high-density snp genotyping microarrays.
\newblock {\em PLoS genetics}, 4(8):e1000167, 2008.

\bibitem{huang2022detecting}
W~Ronny Huang, Steve Chien, Om~Thakkar, and Rajiv Mathews.
\newblock Detecting unintended memorization in language-model-fused asr.
\newblock {\em arXiv preprint arXiv:2204.09606}, 2022.

\bibitem{jagielski2020auditing}
Matthew Jagielski, Jonathan Ullman, and Alina Oprea.
\newblock Auditing differentially private machine learning: How private is
  private sgd?
\newblock {\em Advances in Neural Information Processing Systems},
  33:22205--22216, 2020.

\bibitem{kairouz2015composition}
Peter Kairouz, Sewoong Oh, and Pramod Viswanath.
\newblock The composition theorem for differential privacy.
\newblock In {\em International conference on machine learning}, pages
  1376--1385. PMLR, 2015.

\bibitem{lu2022general}
Fred Lu, Joseph Munoz, Maya Fuchs, Tyler LeBlond, Elliott Zaresky-Williams,
  Edward Raff, Francis Ferraro, and Brian Testa.
\newblock A general framework for auditing differentially private machine
  learning.
\newblock {\em arXiv preprint arXiv:2210.08643}, 2022.

\bibitem{malek2021antipodes}
Mani Malek~Esmaeili, Ilya Mironov, Karthik Prasad, Igor Shilov, and Florian
  Tramer.
\newblock Antipodes of label differential privacy: Pate and alibi.
\newblock {\em Advances in Neural Information Processing Systems},
  34:6934--6945, 2021.

\bibitem{micochallenge}
Microsoft MSRC.
\newblock Mico.
\newblock \url{https://github.com/microsoft/MICO}, 2022.

\bibitem{pillutla2023unleashing}
Krishna Pillutla, Galen Andrew, Peter Kairouz, H~Brendan McMahan, Alina Oprea,
  and Sewoong Oh.
\newblock Unleashing the power of randomization in auditing differentially
  private ml.
\newblock {\em arXiv preprint arXiv:2305.18447}, 2023.

\bibitem{ramaswamy2020training}
Swaroop Ramaswamy, Om~Thakkar, Rajiv Mathews, Galen Andrew, H~Brendan McMahan,
  and Fran{\c{c}}oise Beaufays.
\newblock Training production language models without memorizing user data.
\newblock {\em arXiv preprint arXiv:2009.10031}, 2020.

\bibitem{shokri2017membership}
Reza Shokri, Marco Stronati, Congzheng Song, and Vitaly Shmatikov.
\newblock Membership inference attacks against machine learning models.
\newblock In {\em 2017 IEEE symposium on security and privacy (SP)}, pages
  3--18. IEEE, 2017.

\bibitem{steinke2023privacy}
Thomas Steinke, Milad Nasr, and Matthew Jagielski.
\newblock Privacy auditing with one (1) training run.
\newblock {\em arXiv preprint arXiv:2305.08846}, 2023.

\bibitem{stock2022defending}
Pierre Stock, Igor Shilov, Ilya Mironov, and Alexandre Sablayrolles.
\newblock Defending against reconstruction attacks with r\'enyi differential
  privacy.
\newblock {\em arXiv preprint arXiv:2202.07623}, 2022.

\bibitem{thakkar2020understanding}
Om~Thakkar, Swaroop Ramaswamy, Rajiv Mathews, and Fran{\c{c}}oise Beaufays.
\newblock Understanding unintended memorization in federated learning.
\newblock {\em arXiv preprint arXiv:2006.07490}, 2020.

\bibitem{tramer2022truth}
Florian Tram{\`e}r, Reza Shokri, Ayrton San~Joaquin, Hoang Le, Matthew
  Jagielski, Sanghyun Hong, and Nicholas Carlini.
\newblock Truth serum: Poisoning machine learning models to reveal their
  secrets.
\newblock In {\em Proceedings of the 2022 ACM SIGSAC Conference on Computer and
  Communications Security}, pages 2779--2792, 2022.

\bibitem{watson2021importance}
Lauren Watson, Chuan Guo, Graham Cormode, and Alex Sablayrolles.
\newblock On the importance of difficulty calibration in membership inference
  attacks.
\newblock {\em arXiv preprint arXiv:2111.08440}, 2021.

\bibitem{zanella2020analyzing}
Santiago Zanella-B{\'e}guelin, Lukas Wutschitz, Shruti Tople, Victor R{\"u}hle,
  Andrew Paverd, Olga Ohrimenko, Boris K{\"o}pf, and Marc Brockschmidt.
\newblock Analyzing information leakage of updates to natural language models.
\newblock In {\em Proceedings of the 2020 ACM SIGSAC conference on computer and
  communications security}, pages 363--375, 2020.

\bibitem{zanella2022bayesian}
Santiago Zanella-B{\'e}guelin, Lukas Wutschitz, Shruti Tople, Ahmed Salem,
  Victor R{\"u}hle, Andrew Paverd, Mohammad Naseri, and Boris K{\"o}pf.
\newblock Bayesian estimation of differential privacy.
\newblock {\em arXiv preprint arXiv:2206.05199}, 2022.

\bibitem{zhang2021wide}
Huanyu Zhang, Ilya Mironov, and Meisam Hejazinia.
\newblock Wide network learning with differential privacy.
\newblock {\em arXiv preprint arXiv:2103.01294}, 2021.

\end{thebibliography}
\end{document}